\pdfoutput=1

\documentclass[journal]{vgtc}                % final (journal style)
\usepackage[bookmarks,backref=false,linkcolor=black]{hyperref} %,colorlinks
\hypersetup{
  pdfauthor = {},
  pdftitle = {},
  pdfsubject = {},
  pdfkeywords = {},
  colorlinks=true,
  linkcolor= black,
  citecolor= black,
  pageanchor=true,
  urlcolor = black,
  plainpages = false,
  linktocpage
}

\usepackage{mathptmx}
\usepackage{graphicx}
\usepackage{times}
\usepackage[usenames,dvipsnames,svgnames,table]{xcolor}
\usepackage{mathptmx}
\usepackage{mathtools}
\usepackage{wrapfig}
\usepackage[ruled,linesnumbered]{algorithm2e}

\onlineid{0}

\preprinttext{}

\title{Converting Basic D3 Charts into Reusable Style Templates}

\author{Jonathan Harper and Maneesh Agrawala}
\authorfooter{
\item
 Jonathan Harper is with the University of California, Berkeley. E-mail: jharper@berkeley.edu.
\item
 Maneesh Agrawala is with Stanford University. E-mail: maneesh@cs.stanford.edu
}

\shortauthortitle{Harper \MakeLowercase{\textit{et al.}}: Converting Basic D3 Charts into Reusable Style Templates}
\abstract{ We present a technique for converting a basic D3 chart into
  a reusable style template. Then, given a new data source we can
  apply the style template to generate a chart that depicts the new
  data, but in the style of the template. To construct the style
  template we first deconstruct the input D3 chart to recover its
  underlying structure: the data, the marks and the mappings that
  describe how the marks encode the data. We then rank the perceptual
  effectiveness of the deconstructed mappings. To apply the resulting
  style template to a new data source we first obtain importance ranks
  for each new data field. We then adjust the template mappings to
  depict the source data by matching the most important data fields to
  the most perceptually effective mappings. We show how the style
  templates can be applied to source data in the form of either a data
  table or another D3 chart. While our implementation focuses on
  generating templates for basic chart types (e.g. variants of bar
  charts, line charts, dot plots, scatterplots, etc.), these are the
  most commonly used chart types today.  Users can easily find such
  basic D3 charts on the Web, turn them into templates, and
  immediately see how their own data would look in the visual style
  (e.g. colors, shapes, fonts, etc.) of the templates.  We demonstrate
  the effectiveness of our approach by applying a diverse set of style
  templates to a variety of source datasets.
} % end of abstract

\keywords{Chart restyling, Reusable style templates, Declarative representation, D3 deconstruction, Vega-lite}

\CCScatlist{ % not used in journal version
 \CCScat{K.6.1}{Management of Computing and Information Systems}%
{Project and People Management}{Life Cycle};
 \CCScat{K.7.m}{The Computing Profession}{Miscellaneous}{Ethics}
}

\teaser{
  \centering
  \includegraphics[width=\textwidth]{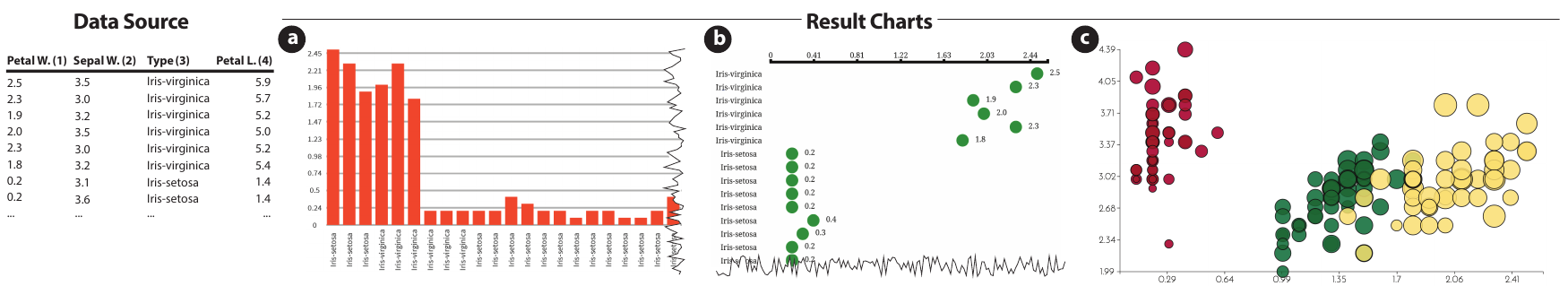}
  \caption{Converting basic D3 charts (e.g. bar charts, line charts, dot plots, scatterplots, etc.) into reusable style templates lets users explore a variety of visual styles for their data. Once we have constructed the templates (see Figure~\ref{fig:results} for the original template charts) the user can specify a new source data table (left) with the importance of each data field (black number in parentheses) and can then apply the template to immediately produce a chart that maintains the visual style (colors, shapes, fonts, etc.) of the template but depicts the source data. Note that due to space limitations we have clipped the right side of the bar chart (a) and bottom of the dot plot~(b).
    {\bf \em (We encourage readers to zoom in to see chart details such as fonts and gridlines.)} %\maneesh{Put full size images in supplementary material?}
  }
\vspace{-0.05in}
  \label{fig:teaser}
}

\begin{document}


\firstsection{Introduction}

\maketitle

Designing visually appealing charts that convey data clearly requires
navigating a large space of visual styles. Designers must carefully
choose visual attributes (e.g. position, size, shape, color, font) for
the data encoding marks (e.g. bars in a bar chart or points in a
scatterplot) as well as the non-data encoding elements (e.g. tick
marks, gridlines, text labels) in the chart.  Although researchers
have developed design principles for making these
choices~\cite{cleveland1985,Mackinlay86,mackinlay2007}, the principles
are not widely known; poorly designed charts that are visually
unappealing and hinder understanding by obscuring the data, are
ubiquitous~\cite{wtfviz}.

Existing visualization tools like Excel, Tableau, Spotfire, and
R/ggplot2 provide a default visual style for the charts they produce.
While these tools usually offer controls for manually tweaking the
visual attributes of the resulting charts, altering the default style
can be tedious. Most users end up exploring a very small region of the
design space centered around the default style and the charts produced
by these tools often look homogeneous.

In contrast, the Web contains a large collection of charts
in a wide variety of different visual styles.  These examples can help
designers better understand the space of possible visual
styles~\cite{herring2009}. Moreover, for novice designers it is often
far easier to select the desired style from a set of examples than it
is to generate a new style from scratch~\cite{lee2010,ritchie2011}.
But existing visualization tools do not provide any support for
re-using the visual style of an example chart.

\begin{figure*}[t!]
  \centering
%  \vspace{-0.1in}
  \includegraphics[width=\textwidth]{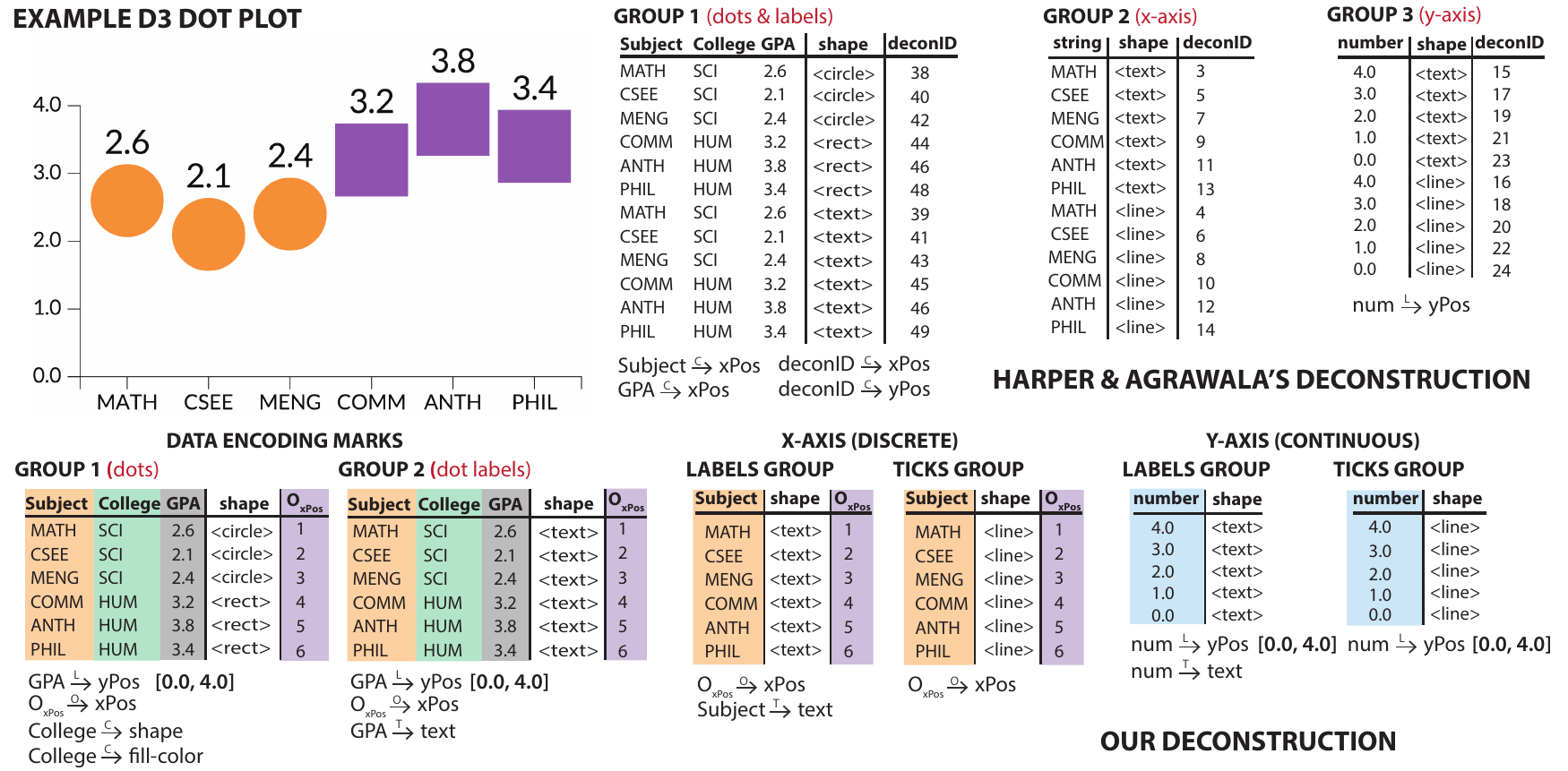}
    \vspace{-0.25in}
  \caption{Harper and Agrawala's deconstruction tool (top) extracts three
    groups of marks for the example dot plot along with five mappings
    (four mappings for Group 1 and one mapping for Group 3). 
We extend their tool to recover much more of the chart's structure (bottom). Our tool
identifies two groups of data encoding marks, as well as a discrete
x-axis and a continuous y-axis, each comprised of two groups of marks
(labels and ticks). We unify data fields across the mark groups as indicated by the corresponding colored backgrounds. We construct attribute ordering $O_{attr}$ fields for each mark attribute (only $O_{xPos}$ is shown).
We recover 13 mappings, including several
attribute order mappings (denoted {\tiny $\xrightarrow{O}$}) and text mappings (denoted~{\tiny $\xrightarrow{T}$}). We also construct a data domain
for each linear mapping (e.g. the data domain for {\em GPA}~{\tiny $\xrightarrow{L}$}~{\em yPos} is [0.0, 4.0]). Note that we have added the red labels in parenthesis to some of the groups to make it easier for readers to match the groups to the chart. These labels are not recovered by either tool.
}
  \label{fig:decon}
%  \vspace{-0.15in}
\end{figure*}

The dominant tool for constructing Web-based charts is the D3
JavaScript library~\cite{bostock2011}. A growing community of D3
developers has already published tens of thousands of D3 charts
online~\cite{bostock,ros,viau}. But despite its widespread use, D3
remains a tool for skilled developers and replacing data or changing
the visual look of an existing D3 chart (e.g. converting a bar chart
into dot plot) usually requires significant re-coding.

Vega~\cite{vega} and Vega-lite~\cite{Wongsuphasawat2015}
have introduced higher-level declarative languages that allow users to
declaratively specify a chart as a collection of mappings between data
and marks. Each mapping describes how a visual attribute of the mark
(e.g. position, size, color etc.) encodes the corresponding data
field. Mackinlay~\cite{Mackinlay86} has shown that this mapping-based
representation significantly reduces the amount of code necessary to
specify a chart while remaining expressive enough to generate a wide
variety of basic chart types (e.g. bar charts, line charts,
dot plots, scatterplots, etc.). Today however, relatively few Vega/Vega-lite
charts are available online compared to D3, and Vega-lite currently
provides only one default visual style for the charts it generates.

Harper and Agrawala~\cite{Harper14} recently developed a technique for
deconstructing existing SVG-based D3 charts to recover their
underlying structure; the data, the marks and the mappings between
them. While they also provide a graphical interface for interactively
restyling D3 charts, their tool is primarily aimed at visualization
experts. All design decisions are left to the user, who must
manually modify the deconstructed mappings to adjust the look of a
chart and their interface does not allow users to replace the
underlying data.

%% In this paper we %take the first step towards such style re-use as we
%% introduce a tool for automatically transferring the visual style from
%% one D3 chart to another. Our tool takes two charts as input, a {\em
%% data source} and a {\em style target} and outputs a new chart that
%% presents the data from the source in the style of the target
%% (Figure~\ref{fig:teaser}).

In this paper we introduce an algorithm for converting a basic D3
chart into a reusable style template. Applying the resulting style
template to a new data source produces a chart that depicts the new
data, but in the visual style of the template chart
(Figure~\ref{fig:teaser}).
%
%Our work is predicated on the assumption that charts typically map
%the most important data to the most perceptually effective mark
%attributes.
%For common charts like bar charts, scatterplots and line charts this
%assumption holds -- the most important data is mapped to
%position or height.
To convert a D3 chart into a style template we first
deconstruct the chart using an extension of Harper and Agrawala's
approach. We then rank the perceptual effectiveness of the
deconstructed mappings based on 
prior work in graphical
perception~\cite{cleveland1985, Mackinlay86}.
To apply the resulting style template to a new data source we first
obtain importance ranks for each new data field.
%Following Mackinlay~\cite{Mackinlay86}, our approach is to map the most
%important data to the most perceptually effective mark attribute.
We then adjust the
template mappings to depict the source data by matching the most
important data fields to the most perceptually effective mappings.

We show how our style templates can be applied to source data in the
form of either a user-specified data table or another basic D3 chart.
%
%While our algorithms are general enough to convert any basic D3 chart
%that can be described as a set of mappings between data and marks,
%into a reusable template,
Our proof-of-concept implementation focuses
on constructing reusable templates for several common chart types:
variants of bar charts, line charts, dot plots, and scatterplots.
%While
%our approach focuses on constructing templates for the most common
%chart types (e.g. bar charts, line charts, scatterplots, etc.)
Users can easily find such basic D3 charts on the Web and immediately
see how their data would look in the visual style (e.g. colors,
shapes, fonts, etc.) of the templates.  We demonstrate that our
templates enable quick exploration of visual chart styles by applying
a diverse set of style templates to a variety of source
datasets. Unlike previous chart design tools, our approach lets users
focus primarily on their data rather than designing the visual
appearance of a chart from scratch or relying on a predefined default
chart style.
%

%Mackinlay~\cite{Mackinlay86}, Harper and Agrawala~\cite{Harper14}, and
%Vega-lite~\cite{Wongsuphasawat2015} have previously shown that the
%declarative mapping-based representation of charts is expressive
%enough for authors to describe a \purple {variety of basic chart types.}
%Our work complements these results and shows that the mapping-based
%representation is high-level enough to allow programmatic manipulation
%of a chart's visual appearance and structure.

%% In this paper we introduce a technique for automatically transferring the
%% visual style from one D3 chart to another (Figure~\ref{fig:teaser}).
%% Given two charts as input, a {\em data source} and a {\em style
%% target}, we first deconstruct each chart using an extension of Harper
%% and Agrawala's technique.
%% %
%% We then analyze the deconstructed representations to rank the
%% importance of the source data fields as well as the perceptual
%% effectiveness of the style target mappings. To build the result chart
%% we update the target mappings with source data fields based on their
%% ranks.
%% %
%% We demonstrate how our tools can be used to transfer style between a
%% wide variety of D3 charts, allowing users to quickly
%% explore the visual design space of charts.

Our contributions include:
\begin{itemize}
\item \textbf{\em Algorithm for constructing style templates from D3 charts.}  We
  extend the approach of Harper and Agrawala to recover additional
  structure from D3 charts, including new types of mappings and
  relationships between data fields.  We demonstrate that the extended
  representation fully captures the structure of many common chart types
  %the common basic chart types
  and can be directly translated into the mapping-based
  representation used by Vega-lite~\cite{Wongsuphasawat2015}.  We
  develop new techniques for ranking the perceptual effectiveness of
  mappings. We show that the additional structure and the rankings are
  crucial for converting D3 charts into reusable style templates.
 \item \textbf{\em Algorithm for applying style templates to new data sources.}  We
   provide an algorithm for applying the resulting style templates to
   any user-specified data table or D3 chart.
   %Our key idea is to match
   %  most important source data fields with the most perceptually
   %  effective template mappings.
   If the new source data is a table we assume the user has specified
   the importance of the data fields. If the new source data is another
   D3 chart we show that we can infer the importance of the
   deconstructed data.
 \item \textbf{\em Evidence for power of mapping-based chart representation.}
 Mackinlay~\cite{Mackinlay86}, Harper and Agrawala~\cite{Harper14}, and
 Vega-lite~\cite{Wongsuphasawat2015} have previously shown that the
 declarative mapping-based representation of charts is expressive
 enough for authors to describe a variety of basic chart types.
 Our work complements these results and shows that the mapping-based
 representation is high-level enough to allow programmatic manipulation
 of a chart's visual appearance and structure.

\end{itemize}

\section{Related Work}
\label{sec:relWork}

Constructing a chart requires mapping data to the visual attributes
(e.g. position, area, color) of graphical
marks~\cite{bertin1983,Mackinlay86}.  While a number of programmatic
chart construction tools such as InfoVis Toolkit~\cite{fekete2004},
%Prefuse~\cite{heer2005}, Protovis~\cite{bostock2009},
ggplot2~\cite{wickham2009}, and Vega/Vega-lite~\cite{vega,Wongsuphasawat2015} have been designed to
facilitate this mapping process, D3~\cite{bostock2011} has become the most
popular Javascript library for producing charts for the Web.   
%Tens of thousands of D3 charts are available
%online~\cite{bostock,ros,viau} 
%A growing community of D3 developers has already published tens of
%thousands of D3 charts online~\cite{bostock,ros,viau}.
%and media websites regularly publish D3 charts that
%reach hundreds of thousands of viewers~\cite{}.  \maneesh{Should reword
%  last parts of paragraph to differentiate from intro to last year's
%  paper.}  
%But despite its widespread use, D3 remains a tool for
%skilled developers and significantly changing the visual look of a D3
%chart (e.g. converting a scatterplot into a bar chart) usually
%requires significant re-coding.
Our work converts existing basic D3 charts into
reusable style templates that can be easily applied to new data
sources.
%so that users can quickly explore the space of
%visual designs rather than focusing on code.

Chart design tools like Excel, Google Spreadsheets,
Polaris/Tableau~\cite{stolte2002}, Lyra~\cite{satyanaryan2014} iVoLVER~\cite{Mendez2016} and Data Driven Guides~\cite{kim2017}
allow users to specify the mappings between data and mark attributes
through a graphical user interface. However, users must rely on their
own expertise to choose the appropriate
mappings.
Mackinlay~\cite{Mackinlay86,mackinlay2007} was the first to develop an
algorithm for constructing basic charts by automatically mapping the
most important data fields (as specified by the user), to the most
perceptually effective mark attributes (as determined via graphical
perception studies~\cite{cleveland1985}).
%Mackinlay~\cite{Mackinlay86,mackinlay2007} was the first to
%build a chart construction tool that could automatically map the most
%important data fields (as specified by the user), to the most
%perceptually effective mark attributes (as determined via graphical
%perception studies~\cite{cleveland1985}).
Our work inverts this
process; given a chart we rank the importance of the
recovered data fields based on perceptual effectiveness of the
attributes they map to.
%We assume that the most important data is
%mapped to the most perceptually effective attribute.

Deconstructing a chart involves recovering its data, its marks and the
mappings that relate them. Researchers
have developed a number of image processing techniques for recovering
marks and data from bitmap images of
charts~\cite{zhou2000,huang2007extraction,yang2006,savva2011}. While
bitmaps are the most commonly available format for charts, accurate
extraction remains challenging because of low image resolution, noise
and compression artifacts.
Despite such inaccuracies recent work has shown that it is possible to
use the recovered marks and data to aid chart reading by adding
graphical overlays~\cite{kong2012} and by connecting the chart to
explanatory text in the surrounding document~\cite{Kong14}.

Harper and Agrawala~\cite{Harper14} focus on deconstructing D3
charts. Their approach recovers the marks and data with 100\% accuracy
and also recovers many of the mappings relating the data to the
marks. Our work builds on their deconstruction approach.  However, we
significantly extend their deconstruction tool to recover additional
chart structure, including new types of mappings and relationships
between data fields. While Harper and Agrawala demonstrate a manual
tool for restyling charts using their deconstructions, we show how the
the additional structure we recover allows us to create style
templates that can be applied directly to new data sources with no
additional user effort.

Our work is inspired by recent techniques for manipulating
  visualizations. Transmogrification~\cite{brosz2013} lets users apply
  user-specified warps to images of charts and thereby produce
  new visual forms. Bigelow et al.~\cite{bigelow2017} develop tools
  that allow users to easily move visualizations between programmatic
  construction tools like D3 and drawing tools like Adobe Illustrator,
  so that they can be edited wherever it is most convenient. Unlike
  these manual tools however, our work focuses on automating the chart styling
  process via reusable style templates.

Style transfer is a well studied problem in Computer Graphics.
Researchers have developed a number of methods for transferring local
characteristics such as texture~\cite{Efros01},
color~\cite{Reinhard01,Pitie05}, non-photorealistic
effects~\cite{Hertzmann01,Ritter06},
%contrast~\cite{Bae06}
and noise~\cite{Chen09,Sunkavalli10}, from one image to another.
These methods rely on non-parametric learning and signal processing
techniques to separate the style of the image from its content and
then apply the resulting model of the style to a new image.  However
these techniques cannot capture higher-level aspects of design
(e.g. fonts, color palettes, line thickness) or domain-specific
semantics (e.g. chart type, axes) and therefore are not well
suited to our problem of transferring style between charts.

Our work is similar in spirit to Bricolage~\cite{Kumar11}, an
example-based tool for restyling webpages. Given two webpages, a
content source and a style target, Bricolage matches page elements
that are visually and semantically similar and then transfers the
content from each source page elements to best matching target page
element. Our style transfer approach similarly considers visual,
perceptual and semantic similarity between elements of two input
charts to determine how to map data from the data source chart to mark
attributes of the style target chart.

%%  LocalWords:  InfoVis Prefuse Protovis ggplot Javascript Mackinlay
%%  LocalWords:  Savva et al ReVision WebPlotDigitizer Agrawala attr
%%  LocalWords:  photorealistic Bricolage webpages Agrawala's xPos
%%  LocalWords:  yPos

\section{Deconstruction}
\label{sec:decon}

Our style templates explicitly represent the visual structure of a
chart as a set of mappings between its data and its visual mark
attributes. 
%Our style templates explicitly represent the visual structure of a
%chart including its data, its visual mark attributes and the mappings
%that relate the data to the attributes.
Harper and
Agrawala~\cite{Harper14} recently developed a tool for deconstructing
basic SVG-based D3 charts into this representation.
While D3 is general enough to work with other Web-based graphics APIs
like Canvas and WebGL, we have found that SVG-based D3 charts are most
common, likely because SVG provides a well-known scene graph
representation for 2D graphics.  To extract the data and associated
mark information their tool uses the fact that D3 charts bind input
data to the SVG nodes representing marks.  Their tool further checks
if there are any linear or categorical mappings (denoted {\em data
  field}~$\rightarrow$~{\em mark attribute}) that explain the
relationships between the data and the mark attributes.  They
represent linear mappings as linear functions that take quantitative
data values to quantitative attribute values. They represent
categorical mappings as a table of correspondences between unique data
values and unique attribute values.

We extract additional structure from D3 charts by extending their
deconstruction tool in six ways:
% We extend their deconstruction tool in six ways to extract additional
% structure from a D3 chart: 
(1) we explicitly label data-encoding marks and axis marks as well as
axis orientation (x- or y-axis) and whether the axis represents
continuous or discrete data, (2) we group marks to identify additional
mappings, (3) we construct data fields (e.g. $O_{xPos}$)
to represent the ordering of the marks with respect to each mark
attribute, (4) we unify data fields which are the same across mark
groups, (5) we identify text format mappings which generate the text
strings for text marks from the values in a data field, and (6) we
compute data domains -- the range of meaningful input values -- for
each linear mapping.  Figure~\ref{fig:decon} shows the additional
structure we recover when we deconstruct a dot plot chart using our
extensions.
We describe these extensions in detail in Appendix~\ref{appendix:extensions}.

Note that even with these extensions our deconstructor is limited to
basic charts -- those that can be described as a collection of
mappings between the data and mark attributes. Our implementation also
inherits a few limitations from Harper and Agrawala and cannot handle
certain types of charts including those that contain non-linear
functional mappings (e.g. log scales), algorithmic layouts
(e.g. treemaps), and non-axis reference marks (e.g. legends). We
detail all of the limitations of our implementation in
Section~\ref{sec:limitations}.

%If we keep sentence in intro Contributions about equivalence to Vega-lite then we should explain how we go directly from Vega-lite to our representation and vice versa. My main question is how we handle axes. Vega-lite probably generates these implicitly. --- Maybe explain that Vega-lite just allows specifications  of data-encoding marks and their mappings. Reference marks like axes and their mappings are generated implicitly. We can directly tanslate our mappings for data encoding marks into Vega-lite encodings  and vice versa. 

\begin{figure}[t]
  \centering
  \vspace{-0.05in}
  \includegraphics[width=\columnwidth]{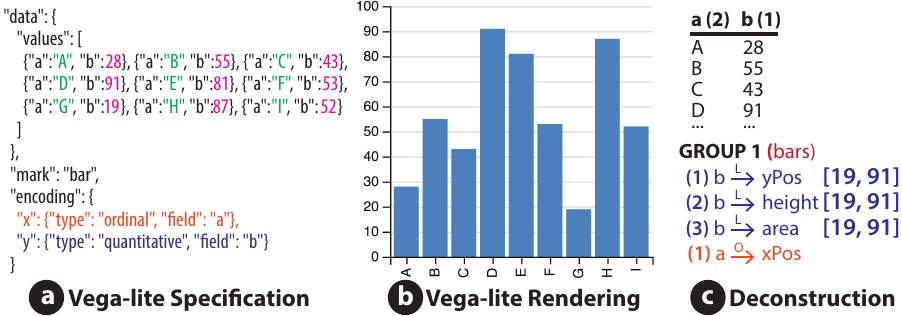}
    \vspace{-0.25in}
    \caption{Vega-lite~\cite{Wongsuphasawat2015} lets users describe a chart as a collection of mappings from data to mark attributes (a). The Vega-lite compiler renders such specifications using a single default visual style (b). Our parser can directly convert Vega-lite specifications into our deconstructed representation by analyzing the Vega-lite data, marks and encodings (c). 
}
  \label{fig:vegalite-parse}
  \vspace{-0.2in}
\end{figure}

\vspace{0.04in}
{\bf \em Similarity to Vega-lite~\cite{Wongsuphasawat2015}.}
Our deconstructed representation for charts is very similar to the
mapping-based representation of Vega-lite. As shown in
Figure~\ref{fig:vegalite-parse}a Vega-lite specification consists of
data, marks and a collection of mappings from the data to mark
attributes. To increase brevity of specification, users do not specify
reference marks (e.g. axes, tick marks, etc.) and their mappings, and
instead rely on Vega-lite to generate them implicitly based on the
data.

But because the representations are so similar, we have developed a
parser that can directly convert such Vega-lite specifications for
data-encoding marks into our deconstructed representation and vice
versa. For example, given the Vega-lite specification in
Figure~\ref{fig:vegalite-parse}a, where the mark type is {\em bar},
our parser directly translates the {\em encodings} into three linear
mappings
{\em b}~{\tiny $\xrightarrow{L}$}~{\em yPos}, {\em b}~{\tiny $\xrightarrow{L}$}~{\em height} and {\em b}~{\tiny $\xrightarrow{L}$}~{\em area} as well as an 
attribute order mapping {\em a}~{\tiny $\xrightarrow{O}$}~{\em xPos}.
%from data field {\em b} to mark attributes {\em yPos}, {\em
%  height} and {\em area}, and an attribute order mapping from data
%field {\em a} to {\em xPos}.
Our parser uses the mark type as well the
the encoding information (e.g. type: {\em quantitative} or {\em ordinal}, field: {\em a}
or {\em b}) to generate these mappings.  It can similarly convert our
mapping-based representation into a Vega-lite specification. Note
however that because Vega-lite does not include reference marks and
mappings we leave those out of these conversions.

%\input{decon-extensions.tex}
%%  LocalWords:  Agrawala SVG Agrawala's attr xPos yPos gridlines
% LocalWords:  rect deconID boolean postfixed postfix metadata versa
%%  LocalWords:  APIs WebGL encodings deconstructor treemaps

%\input{transfer}
%% \begin{figure}[t]
%%   \centering
%%   \includegraphics{./fig/pipeline-2.pdf}
%%   \caption{\harper{Pipeline caption here.}}
%%   \label{fig:pipeline}
%%         % \vspace{-0.15in}
%% \end{figure}

\section{Converting a D3 Chart into a Style Template}
\label{sec:makeStyleTemplate}
%Our approach to templatizing D3 charts relies on the assumption that
%well-designed visualizations map the most important data to the most
%perceptually effective mark attributes.
Deconstructing a basic D3 chart recovers the mappings from the data to
the mark attributes. To convert this deconstructed representation into
a style template we rank the perceptual effectiveness of each
recovered mapping. As we show in
Sections~\ref{sec:applyToDataTable}~and~\ref{sec:applyToD3Chart},
these rankings are essential for applying the resulting style template
to new data sources.

\begin{figure*}[t]
  \centering
  \includegraphics[width=\textwidth]{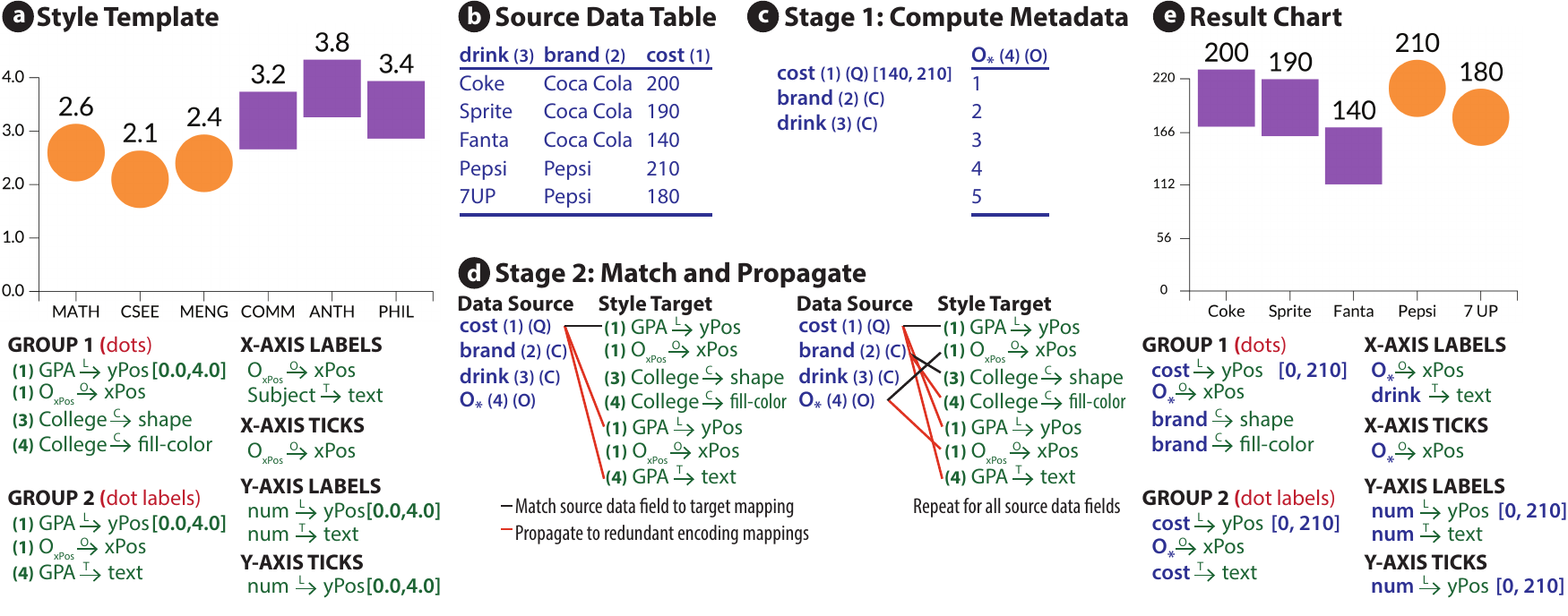}
  \vspace{-0.25in}
  \caption{We apply a style template (a) to a source data table (b) to generate the result chart (e). We convert a D3 chart (depicting average GPA for engineering and humanities colleges) into a style template (a) by deconstructing it and then ranking the perceptual effectiveness of the mappings for the data encoding marks (green numbers in parentheses for mark groups 1 and 2). Given a source data table (containing costs of Coke and Pepsi brand drinks) (b) with importance ranks for each field (blue numbers in parentheses), in stage 1 (c) we compute additional metadata (data type, data domain and ordering field $O_*$). In stage 2 (d) we match the most important data field to the most perceptually effective template mapping, while ensuring that the type of the data field is compatible with the mapping type. Once we find a match we propagate the replacement data field to any other redundant mapping in the template. For example, after matching the source data field {\em cost} to the {\em GPA}~{\tiny $\xrightarrow{L}$}~{\em yPos} mapping we propagate {\em cost} as the replacement for {\em GPA} in every other template mapping, including axis mappings. Finally we generate the marks and rebuild the axes to produce the result chart (e).}
  \label{fig:template-dataTable-result}
  \vspace{-0.1in}
\end{figure*}

\setlength{\columnsep}{0.08in}
\begin{wrapfigure}[13]{r}{1.6in}
\vspace{-0.15in}
\center{
\includegraphics[width=1.58in]{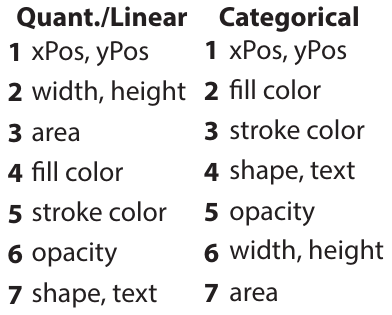}
}
\vspace{-0.25in}
\caption{We adapt Mackinlay's perceptual effectiveness rankings to our set of mark attributes.}
\label{fig:MackinlayRankings}
\end{wrapfigure}
We use Mackinlay's~\cite{Mackinlay86} rankings of the perceptual
effectiveness of visual attributes to set the ranking of each mapping
for the data encoding marks. 
Mackinlay's rankings differ depending
on mapping type 
(quantitative/linear or categorical).
We adapt these
rankings to our set of mark attributes as in
Figure~\ref{fig:MackinlayRankings}. 
Figure~\ref{fig:template-dataTable-result}a 
shows a style template with ranked mappings (numbers in green
parentheses) for the D3 chart from
Figure~\ref{fig:decon}. Note that while we have chosen Mackinlay's rankings for our examples because of their grounding in prior graphical perception research, the rankings are fully customizable within our system.

\section{Applying Style Template to Source Data Table}
\label{sec:applyToDataTable}
Given a style template and a source data table as input, our
goal is to replace the data in the template chart with data from the
source table (Figure~\ref{fig:template-dataTable-result}a,b,e).
%\maneesh{Since well-designed visualizations map the most important data to the
%  most perceptually effective mark attributes}
We also require that an
importance value ($1$ = most important, $N$ = least important) is
associated with each field of the input data source.  In practice we
assume that the user has provided this importance information as part
of the source data table.
For many chart creators this importance information is easy to provide
as they are familiar with the data.
%Our approach allows users to
%remain focused on the data rather than the appearance of the chart
%they are constructing.

We apply the style template to the source data table using a
three-stage algorithm; (1) we first compute additional metadata
(e.g. data type) for the source data table, (2) we then use this
metadata to adjust the mappings for the data encoding marks of the
template chart to depict the source data, and (3) finally we rebuild
the axes of the template chart to serve as reference lines for the
updated data encoding marks.

\subsection{Stage 1: Compute Metadata for Source Data Table}

In the first stage of our algorithm we infer the data type 
(quantitative or categorical) for each field in the source data
table. Specifically, we analyze the data values in each field of the
source table.
%Our algorithm for applying a style template to a source data table
%considers the data type (quantitative or categorical) and data domain
%(range of data values) of each field in the source table. We infer the
%data type by analyzing the data values in each field of the source
%table.
If the field contains only numeric values we set its data
type to {\em quantitative} and if it contains non-numeric values
(e.g. text strings) we set its data type to {\em categorical}.  We
also set the data domain for each quantitative data field to the
min/max range of its data values.

Note that this simple classification heuristic will incorrectly label
numerical categorical data (e.g. employee ID numbers, social security
numbers, etc). as quantitative.  But, because stage two of our algorithm allows
quantitative data fields to serve as input to categorical mappings it can
generate the proper result chart. We
also allow users the option of specifying the data type for any field
in the source table.

%% While this simple heuristic works well in practice, it does classify
%% numerical categorical data as quantitative. For example, a field
%% containing emplyeee ID numbers would be classified as quantitative.
%% However, because our algorithm for applying templates allows
%% quantitative data to serve as input in a categorical mapping
%% (Section~\ref{}) it can properly generate the resulting chart.  We
%% also allow users the option of specifying the data type for any field
%% in the source table.

%Our algorithm also requires an ordering data field for the source data
%table to set the attribute ordering mappings in the style template.
We also extend the source data table to include an {\em ordering}
field $O_*$ that represents the row index of each tuple in the data
table. Since we construct this ordering data field and it does not
represent any of the actual data in the table we assign it
the least importance of all the source data fields. We use this
ordering field in stage two of our algorithm to serve as input to 
attribute order mappings.
Figure~\ref{fig:template-dataTable-result}b,c shows how we analyze
and extend an input source data table.

\subsection{Stage 2: Update Data-Encoding Marks}  

Data-encoding marks are the most important marks in a chart because
their attributes directly encode the underlying data. To update the
data-encoding marks of the template chart so that they reflect the
source data, we first match source data fields to attribute mappings
in the template. For each such match we then synthesize a new mapping
function that maps the source data values to template mark attribute
values.  Finally we generate the marks for the new chart according to
the updated mappings.

\subsubsection{Match Source Data Fields to Template Mappings} 
Algorithm~\ref{alg:matching} describes our procedure for matching the
source data fields to template mappings.
Figure~\ref{fig:template-dataTable-result}d shows our matching
process.
%for the style template and data source table shown in
%Figure~\ref{fig:template-dataTable-result}a,b.
Our approach is to map the most important data to
the most perceptually effective mark attributes.

Our matching algorithm considers each source data field in order by
decreasing importance (line 4) and selects a matching template mapping
based on the data field type (lines 4-24).  Quantitative data can
serve as input to any linear mapping in the style template. Moreover, by
treating each unique numeric data value as a distinct category,
quantitative data can also serve as input to categorical
mappings. Thus, if the data field is quantitative we match against the
top ranked linear or categorical data mapping. A categorical data
field however, may not be numeric and can therefore only serve as
input to categorical mappings in the template. Thus, if the data field
is categorical we match it to the top ranked categorical mapping. Note
that text mappings are treated as categorical mappings in this
matching procedure. Finally we match the {\em ordering} data field
$O_*$ to the top ranked attribute order mapping.

If we find a match we replace the template data field with the source
data field in the matched mapping. The style template may include
redundant encodings in which a single data field maps to several
different visual attributes.  We consider such redundant encodings to
be stylistic constraints and if we replace such a redundantly mapped
data field, we propagate the
replacement to all of the redundant mappings including axis mappings
(lines 25-27 and Figure~\ref{fig:template-dataTable-result}d). 
Once we have completed the propagation we continue the matching
process with the next highest ranked source data field and the
remaining unmatched style template mappings.

If we do not find a match for a source data field our data replacement
result will not depict the data field. Such unmatched data fields
occur when either the data source table includes many more data fields
than the style template chart depicts, or when the source data fields
are incompatible with the template mappings (i.e. the source contains
only categorical data fields, but the style template contains only
linear mappings).
Alternatively, if the style template chart depicts more data fields
than contained in the source data table, some style template mappings
may remain unmatched after one complete pass of the matching loop.  We
can optionally use these extra template mappings to redundantly encode
source data by repeating the matching loop over all of the source data
fields, but only permitting a match with the remaining unmatched 
template mappings. If the style template includes any unmatched
attribute order mappings this redundant encoding approach matches the
ordering field $O_*$ to them.

%% \maneesh{No longer necessary since we always construct $O_*$}
%% If the style template includes attribute order mappings, but the data
%% source does not, the template attribute order mappings will remain
%% unmatched by our algorithm. However, since we construct attribute
%% order data fields for all marks in the data source we match each
%% remaining attribute order mapping of the template to the source
%% attribute order data field with the same attribute name.

%\maneesh{This paragraph only relevant to chart to chart transfer and
%  could probably be cut.}  Finally, note that data fields which are
%not mapped in the data source chart are not considered in our matching
%algorithm and therefore are not depicted in the transfer result.

\begin{algorithm}[t]
\caption{Match source data fields to style target mapping}
\label{alg:matching}
\textbf{Input:} Source data table with importance rank and data type 
for each field. Deconstructed style target chart with perceptual 
effectiveness rankings for each target mapping.\\

\vspace{0.03in}
$U$ = \{target mappings\} \hspace{0.25in} {\em //Set of unmatched target mappings}\\
$M$ = \{\} \hspace{1.04in} {\em //Set of matched target mappings}\\
\vspace{0.03in}
\ForEach{\em source data field $s$ in descending importance order} {
    $match$ = null \hspace{0.53in} {\em //Initialize target mapping match}\\
    \Switch{\em DataType($s$)} {
      \Case{\em Quantitative} {
        $match$ = Top ranked linear or categorical \\
        \hspace{0.1in} mapping in $U$. If tie in rank, pick any one \\
        \hspace{0.1in} of top ranked linear mappings. If no linear \\
        \hspace{0.1in} mapping available, pick any top ranked \\
        \hspace{0.1in} categorical mapping.\\
      }
      \Case{\em Categorical} {
        $match$ = Top ranked categorical mapping in $U$. \\
        \hspace{0.1in} If tie in rank, pick any one of top ranked \\
        \hspace{0.1in} categorical mappings.
      }
      \Case{\em Ordering} {
        $match$ = Top ranked attribute order mapping in \\
        \hspace{0.1in} $U$. If tie in rank, pick any one of top ranked \\
        \hspace{0.1in} attribute order mappings.
      }
    }
    $t$ = DataField($match$) \hspace{0.1in} {\em //Get original data field for match}\\
    DataField($match$) = $s$ \hspace{0.1in} {\em //Set replacement data field}\\
    Propagate $s$ as replacement data field to any other \\
    \hspace{0.1in} mapping in $U$ for which $t$ is the original data field.\\
    Move all such modified mappings from $U$ to $M$.
}
\end{algorithm}

% Rank perceptual effectiveness of each mapping in data source and style target
% Rank importance of each source data field
% Infer data type of each source data field
% Set U = {unmatched target mappings}
% Set M = {} set of matched target mappings
% Loop over source data fields s in descending order of importance
%    Switch DataType(s)
%      Quantitative:
%          match = highest ranked linear or categorical mapping. 
%                   If tie in rank prefer any linear over any categorical,  
%      Categorical:
%          match = highest ranked categorical mapping. 
%                   If tie in rank pick any  
%      Attribute Ordering: 
%          match = highest ranked attribute order mapping. 
%                   If tie in rank pick any  
%    
%    Let t = original target data field of match
%    Set s as replacement data field for match
%    Propagates as replacement data field for any other 
%              mapping in U for which t is the data field 
%    Move all such matched mappings from U to M

\subsubsection{Synthesize Mapping Functions} 
\label{sec:synthmapfunc}
After completing the matching process we synthesize new mapping
functions for each matched pair of source data field and style
template mapping, based on the template mapping's type: linear,
categorical, attribute order, or text.

\vspace{0.05in}
\textbf{\em Linear.}  To synthesize a linear mapping function, we
relate the endpoints of the data domain of the source data field to
the endpoints of mark attribute range for the style template mapping.
Since our matching algorithm ensures that the source data field for a
linear mapping is quantitative we directly look up its data domain as
computed in stage one of our algorithm.
%We look up the data domain for the source data field based on its
%highest ranked linear mapping.
We compute the attribute range of the style template mapping by applying its
mapping function to the endpoints of its original data domain. We then
fit a linear function which maps the start and end points of the
source data domain to the start and end points of the style template attribute
range respectively.  We use the resulting linear function as the
mapping function for our result chart.

Note that for some chart types (e.g. bar charts, dot plots) it is
critical that the chart include the origin at zero when depicting a
quantitative data field. For other chart types (e.g. scatterplots)
including the origin at zero can make it difficult to see the marks.
To handle these two cases we check whether the data domain of the
original template mapping include zero and if so we extend the source
data domain to also include zero. If not, we leave the source data
domain as we computed it in stage one (e.g. the min/max range of the
data values).
%our default policy is to
%check whether the source data domain includes zero and if not we
%extend the source data domain to include zero.
%To handle such cases our default policy is to
%check whether the source data domain includes zero and if not we
%extend the source data domain to include zero.
%
In Figure~\ref{fig:template-dataTable-result}e the linear {\em yPos}
mapping for the dots is synthesized using the source {\em cost} field
with domain [140, 210]. However, the original data domain of the style
template was [0.0, 4.0] and so we extend the data domain of the result
mapping to [0, 210].

\vspace{0.05in}
\textbf{\em Categorical.}  
To synthesize a categorical mapping function, we create a new
correspondence table pairing each unique value of the source data
field with a unique value of the style template mapping's mark
attribute. If the number of unique data values is less than or equal
to the number of unique attribute values our approach generates a
one-to-one correspondence table that serves as the new mapping
function. For example, in Figure~\ref{fig:template-dataTable-result}b the
source {\em brand} data field contains 2 values (Coca Cola and Pepsi) while
the template contains two unique {\em fill-color} values and two unique
{\em shape} values. In this case we assign one fill-color and one shape to each
brand. 
If however, there are more unique data values than unique attribute
values, our approach will leave some of the data values unmatched.  In
such cases we report to the user that it is impossible to construct a
one-to-one categorical mapping and instead we reuse attribute values
in cyclic order so that the same attribute value may be paired with
multiple data values. Although the resulting chart does not correctly
depict the source data -- it visually aggregates distinct data values
-- it can still provide a visual sense for overall look of the
resulting chart.

%% For the {\em color} and {\em shape} attributes we also provide an
%% additional set of predefined attribute values. For {\em color} we
%% include a qualitative color palette from Color
%% Brewer~\cite{brewer}. After our initial pairing process we match any
%% leftover data values with the color from this palette that is furthest
%% away in HSL space from any other color already in the correspondence
%% table. For {\em shape} we include a set of 5 predefined shapes
%% (circle, square, diamond, star, or triangle) and pair leftover data
%% values with any one of these that is not already part of the
%% correspondence table. Although these additonal attribute values make
%% it more likely that our algorithm will generte a one-to-one
%% correspondence table, with enough unique source data values it is
%% still possible for some to remain unmatched.  In such cases we report
%% the issue and again allow the unique attribute values to be paired
%% with multiple data values. \maneesh{Not sure we really need this
%%   paragraph as it is something of a hack.}

\vspace{0.05in}
\textbf{\em Attribute Order.} Attribute order mappings typically
capture information about the chart's layout. To maintain the template
chart's layout while using the ordering data from the source data
table, we update the data field of the template attribute order
mapping, but maintain the linear parameters of the mapping function
unchanged. If the new source field contains more values (or fewer
values) than the template ordering data field this approach extends
(or shrinks) the layout to fit the new number of data values.  The
transfer result in Figure~\ref{fig:template-dataTable-result}e shows an
example of such shrinking as the data source contains only five data
elements while the style template contains six data elements. 

%  \textbf{\em Text.} We treat the function that converts the source data value
%  into a string as the new text mapping function for the result chart.

\vspace{0.05in}
\textbf{\em Text.} For text mappings we create a function that simply
converts the source data value into a string.

\subsubsection{Generate Marks}

%The final step of stage 1 is to apply the newly synthesized mappings to
%generate the data encoding marks for the result chart. 

The final step of stage two is to generate the data encoding marks for
the result chart. For each group of data encoding marks and mappings
in the style template we retrieve the matched data fields from the
source data table. We then join together these data fields into a unified
data table and generate a mark for each row in the resulting table. We set the
attributes for each mark by applying the newly synthesized mappings.
For any unmapped mark attributes, we set the attribute value to the
average (for numeric attributes) or mode (for other attributes) of the
corresponding attribute values from the style template.

%% We finally create the result chart's marks by applying our newly
%% synthesized mapping functions to generate the new chart's mapped mark
%% attributes.  For any mark attributes without a mapping, we set the
%% attribute value for all marks in the new chart to the average (for
%% numeric attributes) or mode (for other attributes) value of the same
%% mark attribute in the style target.

\begin{figure*}[t]
  \centering
  \includegraphics[width=\textwidth]{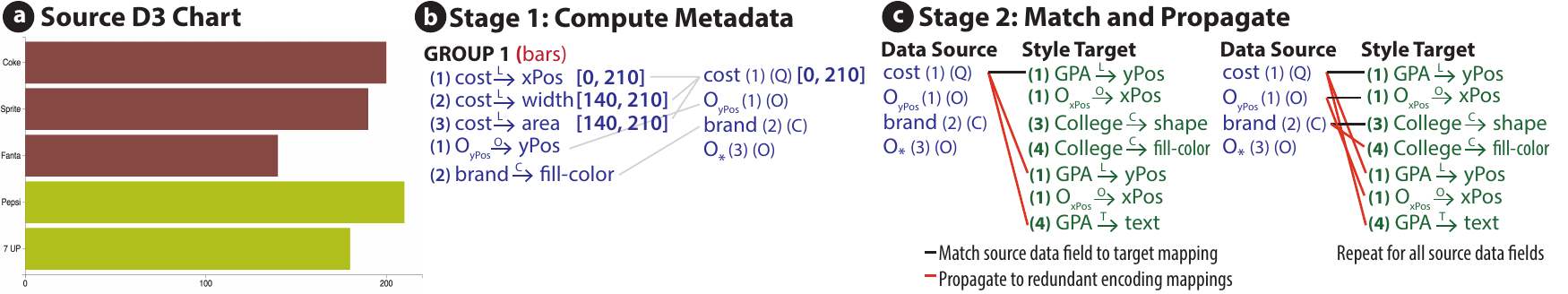}
  \vspace{-0.25in}
  \caption{We can apply style templates to a source D3 chart (a) using the three-stage algorithm of Section~\ref{sec:applyToDataTable} but modifications to stage one (b). We deconstruct the source chart and use the perceptual effectiveness of the mappings for the data encoding marks (b left column) to set the importance of the source data fields (b right column). We also infer the data type, data domain and ordering field $O_*$ using the deconstructed mappings. Finally we apply stages two and three of our algorithm as before to generate a result chart that appear exactly the same as the chart in Figure~\ref{fig:template-dataTable-result}e, but with all attribute order mappings using $O_{yPos}$ from the source D3 chart rather than $O_*$ from the source data table. }
  \label{fig:sourceD3Chart}
%  \vspace{-0.15in}
\end{figure*}

%% \begin{figure*}[t!]
%%   \centering
%%   \includegraphics[width=\textwidth]{./fig/results/results.pdf}
%%   \caption{Our automatic style transfer algorithm applied to a variety of common chart types. The input charts can differ in their mark attribute values (color, fonts, shapes, etc.), data distributions, numbers of marks and even chart type. Our algorithm is robust to these differences and generates result charts that depict the source data in the look of the style target.
%% {\em (Please zoom in to see chart details.)}}
%%   \label{fig:results}
%%         % \vspace{-0.15in}
%% \end{figure*}

\begin{figure*}[t!]
  \centering
  \includegraphics[width=\textwidth]{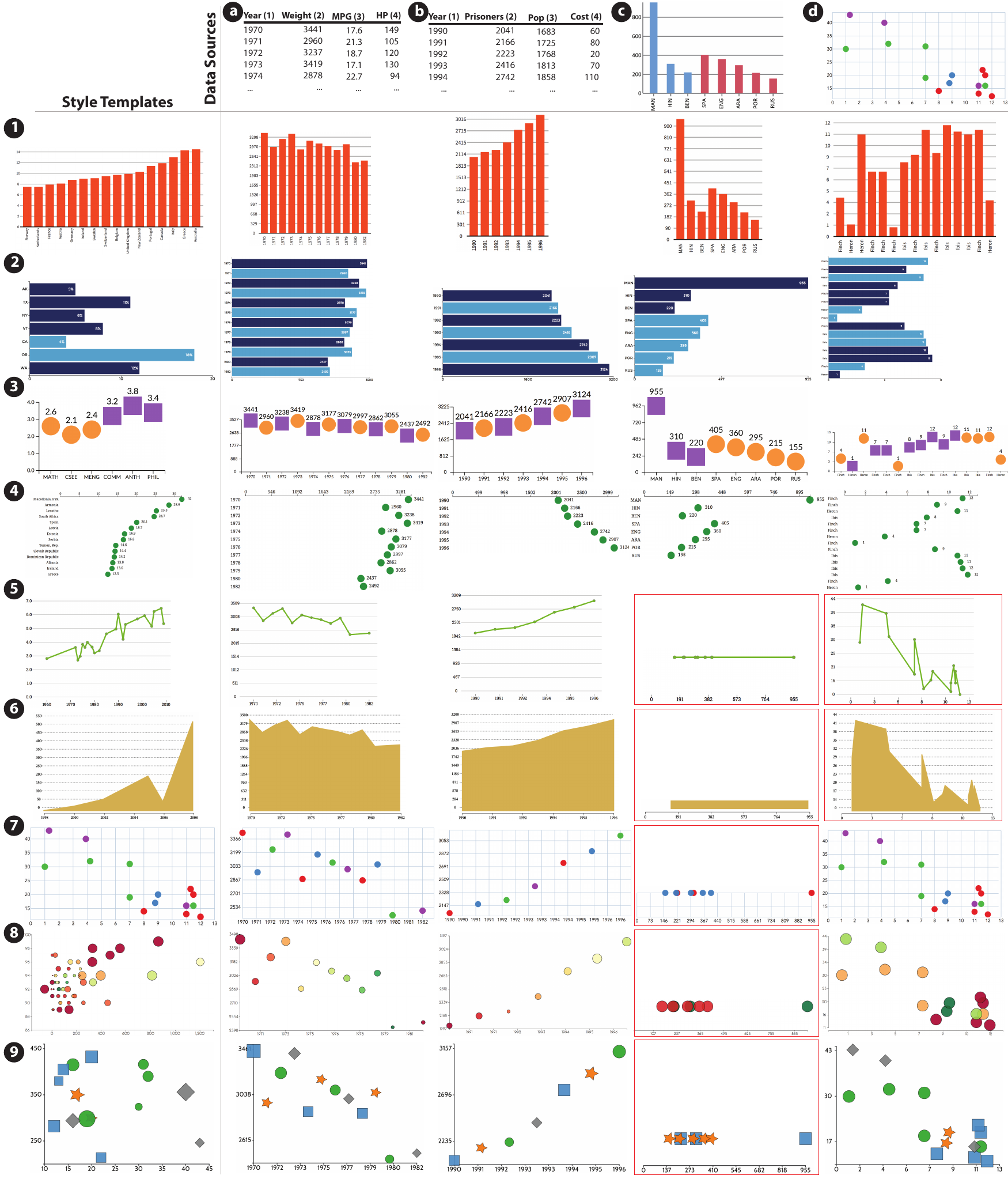}
  \caption{Once we have converted a D3 chart into a style template we can apply it to new data sources in the form of data tables (a,b) or other D3 charts (c,d). The result charts maintain the visual style of the style template chart with similar attribute values (colors, fonts, shapes, etc.) for the marks, axes and labels. Yet, the data values and the numbers of marks differ significantly between the template charts and the result charts. 
Our algorithm is robust to these differences and generates result charts that depict the source data in the look of the style template. Red borders indicate transfer results that our algorithm warns users about as noted in Section~\ref{sec:results}.
{\bf \em (Please zoom in to see chart details like fonts and gridlines.)}}
  \label{fig:results}
        % \vspace{-0.15in}
\end{figure*}

\subsection{Stage 3: Rebuild Axes}

Once we have updated the data encoding marks of the style template, we
rebuild its axes to reflect the new data.  We use a different
rebuilding algorithm depending on whether the axis is continuous or
discrete. We describe the algorithms assuming we are rebuilding an
x-axis; the algorithms for a y-axis are similar.

\vspace{0.05in}
\textbf{\em Continuous axis.}  
A continuous x-axis commonly appears in a scatterplot or a horizontal
bar chart and serves as a reference line relating the positions of the
data encoding marks to data values.  Maintaining the relationship
between the {\em xPos} of the data encoding marks and the {\em xPos}
of the axis tick marks is critical for such an axis to function
properly.

Suppose {\em d}~{\tiny $\xrightarrow{L}$}~{\em xPos} is a linear mapping
from data field $d$ to the {\em xPos} attribute. We can represent the
linear mapping function as a matrix $L$ in homogeneous coordinates where
\begin{equation}
L = \begin{bmatrix}a&b\\0&1\\ \end{bmatrix}
\end{equation}
%% Suppose {\em d}~{\tiny $\xrightarrow{L}$}~{\em xPos} is a linear mapping
%% from data field $d$ to the {\em x-position} attribute. 
%% We use the alternative notation $L\cdot d=$~{\em x-position} where 
%% $L$ is a matrix representing the linear mapping function 
%% $a\cdot d+b$ in homogeneous coordinates as 
%% \begin{equation}
%% L = \begin{bmatrix}a&b\\0&1\\ \end{bmatrix}
%% \end{equation}
and $L\cdot d = a\cdot d+b = $~{\em xPos}. 
Our deconstruction tool recovers the linear parameters $a$ and $b$ for
each such linear mapping.
In this notation we refer to the {\em xPos} mappings for the data
encoding marks and the axis ticks marks in the original style template
as $L_{marks}$ and $L_{ticks}$ respectively.

\setlength{\columnsep}{0.08in}
\begin{wrapfigure}[7]{r}{1.4in}
\vspace{-0.3in}
\center{
\includegraphics[width=1.36in]{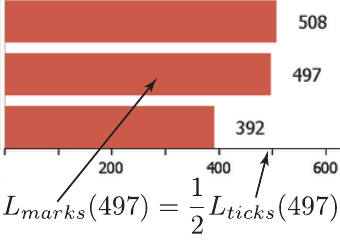}
}
\vspace{-0.05in}
\label{fig:barmappings}
\end{wrapfigure}
Even though the data encoding marks and the x-axis tick marks
depict the same data domain, the mappings 
$L_{marks}$ and $L_{ticks}$ 
may differ. For example, in a horizontal bar chart (inset) the bar
positions are based on the center point of the rectangle and the {\em
  xPos} mapping 
for the bars is $\frac{1}{2}$ the value of the {\em xPos} mapping for the ticks.
Therefore, to rebuild a continuous x-axis for the result
chart we must first recover the relationship between
%the x-position mappings of the data encoding marks and the ticks 
these mappings in the original style template. 

We compute this relationship $R$ between the two mappings as
\begin{align}
R\cdot L_{marks} &= L_{ticks} \\
R &= L_{ticks}\cdot L_{marks}^{-1}
\end{align}
We treat this relationship $R$ between the data-encoding marks and
the axis tick marks as part of the style of the original template chart
that must be maintained when we rebuild the axis for the new data.
After synthesizing a new {\em xPos} mapping for the data
encoding marks in stage two of our algorithm, we denote the new
mapping function as $L_{marks}'$. To build the new x-position mapping
for the axis tick marks $L_{ticks}'$ while preserving the relationship
to the data encoding marks we compute $L_{ticks}' = R\cdot L_{marks}'$
Although this approach updates the mapping from the data to the
x-position of the tick marks, it preserves the positions of the ticks
in image space. It simply changes the data value associated with each
tick mark. To recover these data values we invert $L_{ticks}'$ and
apply it to the tick positions. We then treat these data values as the
data for the text mappings of the axis labels. 
The tick marks of the continuous y-axis in the result chart of
Figure~\ref{fig:template-dataTable-result}e match in position with the
corresponding tick marks in the style template, but the axis labels are
based on the updated source data field.

\vspace{0.05in}
\textbf{\em Discrete axis.}
A discrete x-axis commonly appears in a vertical bar chart or dot plot
and is typically used to label the data encoding marks (e.g. bars or
dots).
%(Figure~\ref{fig:source-target-result} style target and transfer result).
For such charts, our style template contains a unified
attribute order data field $O_{xPos}$, and mappings from this field to
the the {\em xPos} of the ticks, labels and data encoding marks. The
unification ensures when we update the attribute order mapping for the
data encoding marks in stage two, our algorithm will propagate the
update to set the {\em xPos} mappings for the x-axis ticks and
labels. After updating the tick positions we adjust the axis line to
span the new tick marks.

To update the text mappings for the axis labels we search the source
data table for a data field containing a unique value for each row in
the table.
%\maneesh{Maybe cut this paragraph?? It is hard to follow and may confuse more than enlighten. But it is leaving out a detail that is useful for reproducibility.}
%To update the text mappings for the axis labels we consider the source
%data table containing ordering field $O_*$ and search for a data field
%that contains a unique data value for each unique value in $O_*$.
If
our search finds more than one such data field we favor using a
categorical field over a quantitative field as the input for the text
mapping.  If we cannot find any such data field we use $O_*$ as the
data field for the axis label text mapping.  For the example in 
Figure~\ref{fig:template-dataTable-result}e, we find the {\em drink} data
field through this search process and create the {\em drink}~{\tiny
  $\xrightarrow{T}$}~{\em text} mapping for the result chart.

%\maneesh{May want to talk about results for applying to tables here.}

\section{Applying Style Template to Source D3 Chart}
\label{sec:applyToD3Chart}
In some cases users may have access to a basic D3 chart depicting
their data, but wish to quickly explore other chart styles by applying
alternative style templates.  We can apply a style template to a D3
chart using our three-stage algorithm with small
modifications to stage one.
Moreover, our modified algorithm infers the importance values for the
data fields depicted in the source D3 chart and therefore does not
require that the user provide them as part of the input. However the
user can always supply these importances if their preference differs
from the inferred values.
%thereby reducing the amount of work the user must perform.
%
%Unlike the case when applying a template to a source data table, we
%can infer the importance of the data fields depicted in the source D3
%chart and therefore do require the user provide these importance
%ranks.

%\subsection{Changes to Stage 1: Compute Source Metadata}
%Most of the modifications to our template application algorithm occur
%in stage one where we compute medata for the data source.
\vspace{0.05in}
\textbf {\em Computing source data importance ranks.}  We start by
deconstructing the source D3 chart to obtain its data, mark attributes
and mappings.  We focus on the subset of mappings for the
data-encoding marks and assign a perceptual effectiveness ranking to
each one using the same approach we used to construct the style
templates (Section~\ref{sec:makeStyleTemplate}). Based on the
assumption that charts map the most important data to the most
perceptually effective mark attribute, we then directly treat the
effectiveness rank as the importance of the mapped data field.
If the
same data field appears in more than one source mapping we give it the
importance of its highest ranked mapping.
Figure~\ref{fig:sourceD3Chart}a,b shows a source D3 bar chart with
perceptually ranked mappings for the data encoding bars (left column
of~\ref{fig:sourceD3Chart}b), aggregated into importance ranks (right
column of ~\ref{fig:sourceD3Chart}b).

\vspace{0.05in}
\textbf {\em Computing source data types.}
We also use the source mappings to set the data type for each source
data field. If the field is involved in any linear mapping we set its
data type to {\em quantitative} and set its data domain to the domain
of any one of the corresponding linear mappings. Our deconstruction
process ensures that all linear mappings for the same data field are
equivalent.  If a source data field is only involved in categorical or
text mappings we set its data type to {\em categorical}.

\vspace{0.05in}
\textbf {\em Working with source attribute ordering mappings.}  The
deconstructed source D3 chart may also contain attribute order
mappings. We set the data type to {\em ordering} for each data field
involved in such an attribute order mapping.  As in stage one of the
original algorithm we extend the deconstructed source data table with
an ordering data field $O_*$ that holds the row index of each tuple in
the table.  Thus, the deconstructed source D3 chart may provide more
than one ordering data field with different importance ranks
(Figure~\ref{fig:sourceD3Chart}b).  Nevertheless we can apply stage
two of our algorithm without any modification to this deconstructed
source data table (Figure~\ref{fig:sourceD3Chart}c).  In this case
Algorithm~\ref{alg:matching} matches the the most important ordering
data fields of the source chart to the most perceptually effective
attribute order mappings of the style template. Any unmatched
attribute order mapping in the style template is then matched with
$O_*$. In the Figure~\ref{fig:sourceD3Chart} example we first match $O_{yPos}$ and $O_*$ does not need to be used. The result chart for this example appears exactly the same as in Figure~\ref{fig:template-dataTable-result}e, but with all attribute order mappings using $O_{yPos}$ from the source D3 chart rather than $O_*$ from the source data table. 

%In this case we iteratively match the most important
%ordering data field to the most perceptually effective attribute order
%mapping. 

%%  LocalWords:  Agrawala SVG overgroup rect unmapped Mackinlay's
%%  LocalWords:  unranked Mackinlay tuple yPos encodings mapping's
%%  LocalWords:  postfix scatterplot xPos templatizing metadata
%%  LocalWords:  scatterplots gridlines importances

\begin{figure*}[t!]
% \vspace{-0.1in}
  \centering
  \includegraphics[width=\textwidth]{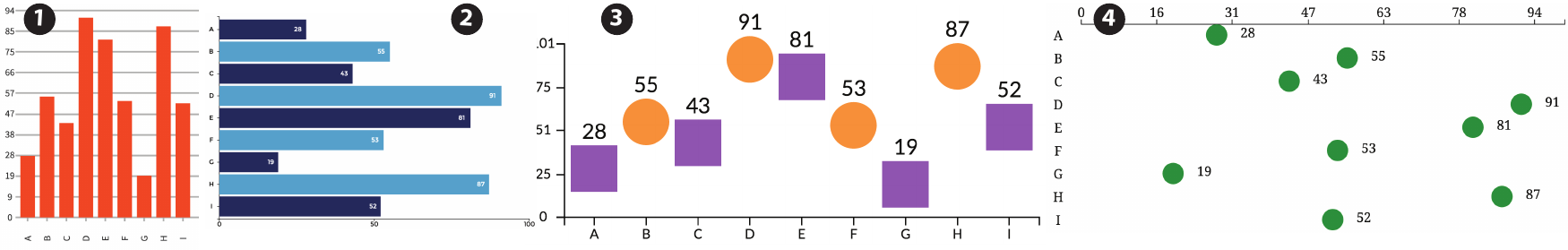}
   \vspace{-0.3in}
  \caption{After deconstructing a Vega-lite~\cite{Wongsuphasawat2015} chart using our parser (original Vega-lite specification and our deconstruction shown in Figure~\ref{fig:vegalite-parse}) we can apply our style templates (original template charts 1-4 shown in Figure~\ref{fig:results} to 
   to this source chart to explore
    additional chart styles that go beyond the Vega-lite default.}
  \label{fig:vegalite-results}
%  \vspace{-0.2in}
\end{figure*}

\section{Results}
\label{sec:results}

We have implemented a pair of tools for (1) converting a
  basic D3 chart into a re-usable style template and (2) applying the
  resulting style template to new source data (either a data table or
  the data in another D3 chart). Our chart conversion tool extends
  Harper and Agrawala's D3 Deconstructor Chrome plugin~\cite{Harper14}
  so that users can click on any basic D3 chart from the Web and
  produce the corresponding style template. Our template application
  tool is a command-line tool that takes a data source file as input
  (either a CSV file with the importance of each field specified as
  metadata, or a deconstructed D3 chart with its corresponding data
  table as produced by the D3 Deconstructor) and produces an SVG-based
  chart that matches the style of the input chart as output. Users can further tweak the resulting chart if necessary using the manual re-styling tools included with the original D3 Deconstructor.

As shown in Figures~\ref{fig:teaser} and~\ref{fig:results}, our
techniques for constructing and applying style templates let users
quickly explore a variety of visual styles for any input source
dataset. When the source data set is given as a data table
(Figures~\ref{fig:teaser} and~\ref{fig:results} cols a,b), the user must
also specify the importance of each data field (numbers in
parentheses). 
When the source data is given as a D3 chart
(Figure~\ref{fig:results} cols c,d), the importance is inferred by our
algorithm.
In these cases users can see how a default chart style (e.g. the Excel
style of the data source in Figure~\ref{fig:results} col c) might be
restyled to produce better looking charts.
All result charts were generated automatically
without any additional user intervention.

The result charts maintain the visual style of the style template
with similar attribute values (colors, fonts, shapes, etc.) for
the marks, axes and labels. Yet, the data values and the numbers of
marks differ significantly between the template charts and the
result charts.  Our algorithm is robust to these differences and
generates result charts that depict the source data with the look of the
style template.
%Our approach ensures that the result charts match the
%type (bar chart, dot plot, et.) and style of the template while
%depicting the source data.
%We discuss a few of the result charts in detail.

The fonts, colors and gridlines vary considerably between the
templates.  A few of the templates include text labels on the marks
(Figure~\ref{fig:results} rows 2,3,4) that redundantly encode data
values to make it easier for viewers to read the exact values.  The
bar charts (Figure~\ref{fig:results} rows 1,2), dot plots
(Figure~\ref{fig:results} rows 3,4), line chart (Figure~\ref{fig:results}
row 5) and area chart (Figure~\ref{fig:results} row 6) 
all include a quantitative axis that starts at zero
because zero is included in the template chart data domain.  In
contrast the scatterplots (Figure~\ref{fig:results} rows 7,8,9) do not
always include an origin at zero.

The blue horizontal bar chart (Figure~\ref{fig:results} row 2) and
orange purple dot plot (Figure~\ref{fig:results} row 3) style
templates use color to depict a two-valued categorical
variable. Several of the result charts for these templates
(Figure~\ref{fig:results} rows 2,3 cols a,b,d) cycle between the
colors because the source data fields that map to color contain more
than two unique values (Section~\ref{sec:synthmapfunc}).  However, the
source data in Figure~\ref{fig:results} col c includes a field with only
two unique data values and the result charts use color to depict it
accurately.

The source data chart in Figure~\ref{fig:results} col c includes only
one quantitative data field. Thus, when we apply the line, area, and
scatterplot chart templates (Figure~\ref{fig:results} rows 5-9) to
this source we produce a 1-dimensional result where the quantitative
data field is mapped to {\em xPos} and the {\em yPos} remains unmapped
so that all of the marks appear on the same horizontal line. While the
line and area charts are difficult to read because of self-occlusions
in this case, the 1-dimensional scatterplots can be useful plots for
seeing the distributions of the quantitative data field.  In all of
these cases the resulting charts are correct in the sense that they
depict the single quantitative dimension of the data source on a
single axis.  The resulting charts also remain visually similar to the
style templates and thereby convey how the data would look using the
template.  However the line and area charts in particular are
difficult to read and therefore whenever we apply a template that has
a mismatch in the number of quantities or categorical mappings from
the number of such fields in the data source we report the mismatch to
the user and suggest that it would be better to pick an alternative
style template that matches the data source. By showing the result
rather along with the warning, the user can see what happens when
there is a mismatch and potentially learn from it.

The source data chart in Figure~\ref{fig:results} col d contains two
quantitative data fields, but neither are monotonic. Thus, when we
apply the line and area chart templates (rows 5 and 6) to this data we
produce charts that can be difficult to read and interpret.  However,
in the line chart case we produce a valid connected
scatterplot~\cite{Haroz2015} where the ordering of the connections is
based on an ordering data field. In all such cases we report the lack
of monotonicity whenever a line or area chart template produces such a
result. Here again, our warning can help users learn what happens when
data points are connected in a non-monotonic order in a line or area
chart.

%% \maneesh{MAY NEED UPDATE: Bring in Vega-lite parsing earlier in paper. Explain how we can go beyond the single default Vega-lite style.}
%% Finally, we note that our deconstructed representation of charts is
%% very similar to Vega-lite~\cite{Wongsuphasawat2015} -- a specification
%% language in which a chart is described as a collection of mappings
%% from data fields to mark attributes (Figure~\ref{fig:vegalite}a). The
%% Vega-lite compiler takes such specifications and renders them into
%% charts using a single default visual style (with pre-selected fonts,
%% colors, etc.). We have developed a parser that can directly convert
%% such Vega-lite specifications of charts into our deconstructed
%% representation and vice versa.
%%   %Note that while
%%   %Vega-lite also allows specification of data transformations
%%   %(e.g. aggregations), our parser ignores such these transformations.
%% Using our deconstructor and parser tools we can convert D3 charts into
%% Vega-lite, so that users can apply the Vega-lite toolchain to D3
%% visualizations~\cite{vega}.  Moreover users can apply our style
%% templates to Vega-lite charts to go beyond the default Vega-lite
%% visual style (Figure~\ref{fig:vegalite}d).

\vspace{0.05in}
{\bf \em Stylization tool for Vega-lite~\cite{Wongsuphasawat2015}.}
As noted in Section~\ref{sec:decon}, we have also developed a parser
for converting Vega-lite specifications of common chart types into our
deconstructed representation and vice versa. Vega-lite currently
offers a single default visual style (with pre-selected, fonts,
colors, axis thicknesses, absence of gridlines, etc.). With our parser
and style transfer approach we can take a Vega-lite specification as
input (Figure~\ref{fig:vegalite-parse}), and apply any template D3 charts we have generated to try out
different appearances for the chart (Figure~\ref{fig:vegalite-results}).
In this case since we are applying the templates to a source chart
rather than a data table we can use the automatic approaches for
computing data importance ranks, source data types and also make use of
the source attribute ordering mappings (Section~\ref{sec:applyToD3Chart}).  Similarly we can apply our
deconstructor to common D3 charts, parse the resulting deconstruction
into Vega-lite and apply the Vega-lite toolchain to the D3
visualization~\cite{vega}.  

\vspace{0.05in}
       {\bf \em User Feedback.}
To further understand the usefulness of our tools we showed them to
seven professional data analysts who visited our lab for a
visualization workshop as well as three professional journalists
experienced in chartmaking.  The data analysts were familiar with
tools like Excel and R/ggplot2 but did not go much beyond the defaults
in generating charts.  The journalists regularly used a variety of
chart construction software including Excel, Adobe Illustrator,
Tableau as well as programmatic tools like R/ggplot2 and D3. Unlike the data analysts they were experts in these construction tools.

After a brief introduction explaining the capabilities of our tools,
we showed them how our tool could be used to take any data table,
specify the importance of each field and immediately see the data in a
variety of styles, using a set of 15 templates we had constructed
earlier. We offered to let the visitors try stylization on their own
datasets and four of them took us up on this offer. One of the
journalists who primarily worked in D3 also asked us to convert one of
his own D3 bar charts into a style template and tested the template on
several of our datasets.

While the visitors gave us oral feedback throughout the demonstrations
we explicitly asked them to provide qualitative feedback about our
tools via a written feedback form right before they left. On a 5 point
Likert scale ranging from strongly disagree to strongly agree, all of
the visitors wrote that they agreed or strongly agreed that
``Re-usable style templates make it quick and easy to see data in a variety
of styles''.  They also agreed or strongly agreed that ``Choosing a
basic D3 chart from the Web as a style template is useful.'' Note that
only one of the visitors tested the ability to use a D3 chart as a
template, but he told us that he was satified with the results he
obtained when applying the template to new data sets.

The experienced journalists did mention that a few of our results
could still use a bit of tweaking (e.g. spacing gridlines, reorienting
text labels, etc.) before they would be ready for publication.  They
were happy to learn that they could use the manual re-styling toools
of Harper and Agrawala's D3 Deconstructor~\cite{Harper14} or load the
resulting chart into an SVG editor like Adobe Illustrator to perform
such tweaks.  Overall, they thought that the stylized charts produced
by applying our templates were excellent starting points and could
save them hours of time in the initial chart design stage.

While this qualititative user feedback suggests that our re-usable
style templated offer useful functionality to both novice and expert
users, we believe that a formal user study is an excellent direction
for future work in order to fully evaluate the effectiveness of our
tools.

\subsection{Limitations}
\label{sec:limitations}
Although our technique for applying style templates successfully handles a
variety of input charts it does have some limitations.
Our approach requires a structural representation of the data, marks
and mappings of a chart. While our deconstruction tool can produce
this representation for SVG-based D3 charts, it is currently limited to basic
chart types (e.g. variants of bar charts, scatterplots, dot plots,
line charts, etc.) with linear, categorical, attribute order and text
mappings. Specific limitations include:

\vspace{0.05in}
\textbf{\em Cannot recover non-linear functional mappings.} Our
  deconstructor cannot recover non-linear functional mappings
  (e.g. logarithmic, polynomial exponential, etc.) between the data
  and mark attributes. Extending our deconstructor to use 
  function fitting techniques to test whether commonly used mappings
  functions (e.g. log scales) produce good fits whenever the linear
  mapping cannot be generated is a direct next step for our work.

\vspace{0.05in}
\textbf{\em Cannot fully manipulate mark shape.} Like Harper and
  Agrawala's~\cite{Harper14} deconstructor, we parameterize the
  geometric attributes of marks using their bounding boxes. While this
  approach lets us recover mappings from data to many geometric mark
  attributes including position, x-scale, y-scale, and area, we cannot
  recover or manipulate mappings to the internal angle of a
  shape. More specifically, given a D3 pie chart our deconstructor
  cannot recover or modify the mapping between the data and the pie
  slice angle.  Modifying our deconstructor to appropriately
  parameterize commonly used mark types such as pie slices is a direct
  extension of our work. Note that Vega-lite also does not support
  generation of pie charts, but if it did then we could apply our
  stylization tool for Vega-lite (see Section~\ref{sec:results}) to
  generate and apply pie chart templates.

\vspace{0.05in}
\textbf{\em Cannot recover algorithmic mappings.}  Some chart
  types like treemaps, jittered scatterplots and force-directed
  node-link graphs use complex algorithmic techniques to choose the
  position of marks. Our deconstructor cannot correctly recover the
  mapping algorithm between data the mark position attribute for these
  charts. Perhaps using more sophisticated program slicing and
  analysis techniques it would be possible to directly extract the
  code implementing the mapping function from the template D3 chart.

\vspace{0.05in}
\textbf{\em Cannot handle non-axis reference marks.} As noted in
  Section~\ref{sec:axismarks} non-axis reference marks such as legends
  are treated as data-encoding marks by our deconstructor breaks our
  style transfer process. Developing techniques for identifying
  such non-axis reference marks and deconstructing them separately from
  the main chart and its axes is an open direction for future work.

\vspace{0.05in}
\textbf{\em Cannot handle interaction and animation.} Our style
  templates focus on capturing the visual appearance and structure of
  basic D3 charts that are static. While some D3 charts include
  interaction and/or animation our techniques cannot capture these
  dynamic aspects of the charts and therefore cannot apply them to new
  data sources. One challenge is to develop a declarative
  representation for interaction and animation. Recent work by
  Satyanarayan et al.~\cite{Satyanarayan17} extends the declarative language of
  Vega-lite to represent certain kinds of chart
  interactions. Converting an interactive D3 chart into this Vega-lite
  interaction specification is an exciting direction for future work.

\vspace{0.05in}
As we have noted the first two of these limitations require extending
the implementation of our deconstructor in relatively direct ways and
would not affect our algorithms for capturing and applying style
templates.  The other limitations are deeper challenges that may
require new algorithmic techniques.  Nevertheless, despite these
limitations, our work shows that our deconstruction tool fully
captures the structure of many of the most common types of basic
charts.

\section{Conclusion}

%% We have presented a technique for converting a D3 chart into a style
%% template and a three-stage algorithm for applying the resulting
%% template to new data sources.  Our approach allows users to quickly
%% try new looks for their charts. In practice we have found that it is
%% significantly faster to apply our style templates than it is to create
%% a chart or restyle an existing chart using common chart-making tools
%% like Excel, Tableau, Illustrator, or even D3.
%% \maneesh{Could move last part to a discussion section.}
%% \harper{Or just integrate higher into this paragraph?}
%% Our techniques operate
%% on a high-level structural representation of charts. Our work shows
%% that this representation is sufficient to capture both the style and
%% content of a chart, and that it can be recovered by analyzing only the
%% data and marks in a chart.

We have presented a technique for converting a basic D3 chart
into a style template and a three-stage algorithm for applying the
resulting template to new data sources. Our algorithm operates on a
high-level structural representation of charts. Our work shows that
this representation is sufficient to capture both the style and
content of a chart, and that it can be recovered by analyzing only the
data and marks in a chart.  Our approach let users quickly try new
looks for their charts. In practice we have seen that it is much
faster to apply our style templates than it is to create a chart or
restyle an existing chart using common chart-making tools like Excel,
Tableau, Illustrator, or D3.

\section*{Acknowledgements}
This work was supported by an Allen Distinguished Investigator Award.

\appendix
\section{Appendix: Extended Deconstruction}
\label{appendix:extensions}
We detail the six extensions we make to Harper and Agrawala's~\cite{Harper14}
D3 chart deconstruction technique to recover additional chart structure. 
\subsection{Label Data-Encoding Marks and Axis Marks}
\label{sec:axismarks}
%% A chart is often composed of two types of marks -- marks that
%% encode the data of the chart, and marks that form axes.  Harper and
%% Agrawala's deconstruction tool does not differentiate between these
%% two types of marks.  Yet, axes are specialized marks because they
%% serve as reference structures that allow viewers to read off the
%% numerical values of the primary data-encoding marks.

A chart is often composed of two types of marks -- {\em data-encoding marks}
that depict the data via their visual attributes, and {\em reference marks},
such as axes and legends,
that
%%serve as reference structures and 
allow viewers to associate the visual attributes of the data-encoding marks 
(e.g. {\em xPos}, {\em yPos}, etc.) with specific data values.
Maintaining the relationships between these two types of marks is
critical for viewers to correctly read the data from a chart.
However, Harper and Agrawala's deconstruction tool does not
differentiate between these two mark types.

We extend their deconstruction tool to explicitly label {\em axis marks},
which are the most common reference marks. We use
the fact that D3 groups together all of the marks comprising an axis
and stores a specialized axis {\em scale object} with the group. In
deconstruction we check whether each SVG group node has an associated
scale object and if so we label all of its child SVG nodes as axis
marks. We also recover two additional properties from the scale
object; the orientation of the axis (x-axis or y-axis) and whether the
axis is a reference for discrete data (e.g. dot plot x-axis) or for continuous data (e.g. dot plot y-axis).
Finally, we examine the geometry
of the axis marks to differentiate tick marks, text labels and the
axis line.  Note that in D3 gridlines are typically created as chart-spanning tick marks,
so we label gridlines as tick marks.
%
%\harper{Need to mention what happens with axis lines.  Need to mention
%  something about different types of axes: continuous vs. ordinal.}
%\maneesh{I tried to address this in second to last sentence.}

After labeling all of the axis marks we label the remaining marks as
data-encoding.  Note that our approach labels non-axis reference
marks, such as legends, as data-encoding marks. This incorrect
labeling can break our style transfer process. %(see Limitations Section). 
We leave it to future work to automatically label such
non-axis reference marks. 

%\maneesh{make sure we address this
%  limitation, or remove the note from this paragraph.}

%% Axes are special parts of a chart because they provide reference
%% structures to relate the data values shown by other marks in the
%% chart.  We extend deconstruction to label the marks corresponding to
%% the axes in a chart so that we can ensure that the relationship
%% between the chart's data-encoding marks and axes.
%% %
%% %To find the marks for each axis, 
%% We use the fact that D3 stores a distinctive D3 scale object to 
%% the top-level SVG group node containing 
%% %all of the marks for 
%% an axis to label its marks as part of the axis.
%% We also check the orientation of the line and tick marks for each axis to determine whether it
%% is an $x$ or $y$-axis.  
%% We use this axis information to ensure that the labels and ticks for 
%% an axis are grouped together (and not grouped with marks from another axis).

%% D3 axes may additionally include special data-specific formatting for axis labels.
%% We use the fact that D3 stores this formatting function with the D3 scale object 
%% bound to the top-level SVG node for the axis to recover the label format.

\subsection{Regroup Marks to Identify Additional Mappings}

In order to construct mappings Harper and Agrawala~\cite{Harper14}
first group together marks that have the same data schema. For each
such group they then identify any linear or categorical functions that
relate the data to the mark attributes. % for all marks in the group.
But this approach can over-group marks and fail to find some of the
mappings.  
%% \maneesh{Maybe put remaining sentences after we describe what we do.}
%% In our example dot plot (Figure~\ref{fig:decon}), the {\tt
%%   <circle>} and {\tt <rect>} marks corresponding to the dots as well
%% as the {\tt <text>} marks corresponding to their labels all share the
%% same data schema and are grouped together by Harper and Agrawala.  As
%% a result they cannot find a mapping between the data and the
%% y-position attribute of the dots or the labels.
We instead start by only grouping together the marks
if they have the same SVG node type (e.g. \texttt{<circle>}, \texttt{<rect>}, \texttt{<text>}). We then construct mappings for each group
independently. 
Finally, for each pair of groups we check whether the mappings are 
equivalent; that is, for each mapping in one group we check whether there is a
mapping in the second group for which the data field and mark
attribute match.
%% Finally, for each pair of groups we check whether the mappings are
%% equivalent; that is, for each mapping we first check that the data
%% field and mark attribute are the same. Then, for linear mappings we
%% check that the linear function parameters are exactly the same, while
%% for categorical mappings we check that the same data values map to the
%% same attribute values. 
If we find a match for all the mappings in both groups, we merge the two
groups and test whether we can construct additional mappings.

For our example dot plot (Figure~\ref{fig:decon}), Harper and Agrawala
group together the \texttt{<circle>} and \texttt{<rect>} marks
corresponding to the dots with the \texttt{<text>} marks corresponding
to their labels because they all share the same data schema. This
grouping prevents their technique from recovering the {\em GPA}~{\tiny
  $\xrightarrow{L}$}~{\em yPos} mapping since the same {\em GPA} value
maps to two different {\em yPos} values (one for the dot and one for
the text) and they instead recover a less useful categorical mapping {\em GPA}~{\tiny
  $\xrightarrow{C}$}~{\em xPos}.  In contrast, our procedure only groups together the
\texttt{<circle>} and \texttt{<rect>} marks while leaving the
\texttt{<text>} marks in a separate group.  It can then recover a {\em
  GPA}~{\tiny $\xrightarrow{L}$}~{\em yPos} mapping for both of these
groups as well as a {\em College}~{\tiny $\xrightarrow{C}$}~{\em
  shape} mapping for the dots.
%\maneesh{Figure should show lack of equivalence in y-position
%   mapping of dots vs. labels}

%% We created this chart using one linear mapping between the value
%% data field and the y-positions of the points and a different linear
%% mapping between the value data field and the y-positions of the
%% labels. But because the marks are grouped together it is impossible
%% for Harper and Agrawala’s approach to construct a single linear
%% mapping describing the relationship between the value data field
%% and the y-positions of the points and labels.  Instead their
%% approach finds no mapping to the $y$-position of the marks,
%% requiring the user to manually split the marks into subgroups to
%% recover the $y$-position mapping.  % they construct a more complex
%% categorical mapping from the % value data field to the y-position
%% of the point and label marks.  % \maneesh{possibly replace prev
%% sentence with the following sentence or % kill these sentences all
%% together (depends on what we show in % figure)} They require the
%% user to manually split the marks into % subgroups in order to
%% recover the proper set of linear mappings.

\subsection{Construct Mark Attribute Ordering Data and Mappings}

In some charts mark attributes are not related to any data field but
instead form a regular ordered sequence in attribute space. For
example, the {\em xPos} of each dot in our dot plot is regularly
spaced in the image. Harper and Agrawala attempt to
recover such ordering information from the SVG rendering order of the
marks in the chart. However, the rendering order does not always
correspond to the attribute ordering and in such cases their approach
will fail to find the ordering.

Our approach for recovering such ordering information is to construct
an {\em attribute ordering} data field $O_{attr}$ for each mark
attribute. We set the data for this field as the sort ordering index
of the corresponding attribute values. For an attribute that is
regularly spaced, we can then recover a linear mapping between this
ordering data and the attribute values to capture the regular
spacing relationship. We call such mappings {\em attribute order 
  mappings}. 
%because the attribute ordering data was constructed by our system
%rather than recovered from the chart.

%However, if we also identify a linear mapping from non-derived data to 
%the same attribute we drop the derived mapping.
%However, if we identify both a derived mapping as well as a non-derived linear mapping 
%to the sane attribute
%If we recover a derived mapping to an attribute which is also mapped
%from non-derived data, we prefer the non-derived mapping as
%non-derived mappings explain the data bound to marks in the chart.
% since the data bound to marks is semantically meaningful in the chart whereas attribute ordering data is constructed by our system.
% However, if we find a linear mapping between any non-derived data field and the attribute we 
In our dot plot example (Figure~\ref{fig:decon}), Harper and Agrawala
only recover a categorical mapping between the SVG rendering order
index {\em deconID} and the {\em xPos} attribute which does not represent the regular spacing between the marks.
%Such a categorical mapping cannot represent the regular spacing
%between the marks. 
In contrast, our approach recovers an attribute
order mapping $O_{xPos}$~{\tiny $\xrightarrow{O}$}~{\em xPos}
that captures the regular spacing as a linear
function. 

%% In some cases a chart's marks are placed based on some ordering on
%% the data rather than the data values themselves.  In our example
%% dot pot in Figure~\ref{fig:decon}, the $x$-position of each dot is
%% determined in this way.  Harper and Agrawala~\cite{Harper14}
%% recover these mappings by storing data about the location of each
%% mark in the chart's SVG tree, which they call its ``deconID''.
%% However, this approach will fail to recover such mappings from
%% orderings when the location of the marks in the SVG tree don't
%% have a linear relationship to their attribute values.  Our
%% approach is to instead check for regularly-spaced mark attribute
%% values by generating a sort ordering data field for each numeric
%% mark attribute and checking for linear mappings from each sort
%% ordering to the mark attribute it was computed from.  We call such
%% mappings on sort orderings ``derived'' mappings.  % compute sort
%% orderings for each numeric mark attribute, and test them for The
%% dot plot in Figure~\ref{fig:decon} displays an advantage of our
%% approach: Harper and Agrawala's deconstruction of the example dot
%% plot finds only a one-to-one mapping from the ``deconID'' field to
%% the $x$-position of the dots whereas our approach finds a simpler
%% linear relationship from the $y$-position ordering.

\subsection{Unify Replicated Data Fields}

Some charts replicate the same data for two different groups of
marks. In our dot plot example (Figure~\ref{fig:decon}), dots and dot
labels replicate the {\em GPA} data field and linearly map these
fields to the {\em yPos} attribute in slightly different ways.  We
unify such replicated data fields so that any change to the data field
propagates to all of the mark attributes mapped by it.  So, if the
unified {\em GPA} data is changed the {\em yPos} for both the
corresponding dot and dot label will be updated.

Our unification approach considers each pair of data fields in the
chart and checks for two conditions: (1) the data field names match
and (2) they contain the same data values, including repeated values
(i.e. we sort the data and check that corresponding values match).
Some D3 charts do not include the data field name with the data bound
to the marks. In these cases Harper and Agrawala generate a data field
name based on the the type of the data (number, string, or boolean).
Since these data values do not have a semantic field name, we attempt
to unify them with all other data fields by only performing the second
check and matching the sorted data values.  
In our example dot plot (Figure~\ref{fig:decon}) we unify a number of
fields across the different groups of data encoding and axis marks.
\subsection{Extract Text Format Mappings}
Charts commonly use text marks to display specific data values.  For
example, our dot plot (Figure~\ref{fig:decon}) labels each dot with a
text mark that shows the exact {\em GPA} value for the dot.  Given the
group of label marks, Harper and Agrawala would recover a categorical
mapping from the {\em GPA} data field to the {\em text} attribute of
the text mark. However, because a categorical mapping is represented
as a table of correspondences between unique data values and unique
attribute values, it is not extensible and cannot convert new data
values into attribute values.  It does not capture the general
functional relationship between the data and the text string
attribute.

We recover a more general and extensible {\em text
  format mapping}. For each categorical mapping between a
data field and a text string attribute we further check if the
string version of the data value matches the corresponding {\em text}
attribute value. If so we set the text format mapping function
to simply convert the data value to a string -- i.e. string(data
value). However, in some cases the data value may be related to the
text string attribute value by a more complicated formatting
function. For example, data representing U.S. dollars may be prefixed
with the ``\$'' symbol or postfixed with the string ``dollars'' when
displayed as a text mark. In such cases we apply a regular expression
parser on the group of text marks to recover the common prefix
and/or postfix and set the text format mapping function to: 
{\em prefix} + {\em string(data~ value)} + {\em postfix}.

\subsection{Recover Data Domain of Linear Mappings}

Although a linear mapping function can be applied to any numeric input
to produce an attribute value, in the context of the chart, only a
limited domain of input data values produce meaningful mark attribute
values. In our dot plot (Figure~\ref{fig:decon}) the {\em GPA}~{\tiny
  $\xrightarrow{L}$}~{\em yPos} mapping for the dots is only meaningful
over the data domain $[0.0, 4.0]$, the limits of the y-axis. We
construct such data domains for linear mappings as follows.

For each linear mapping we first set its data domain to the min/max
range of its data values. However, the resulting domain may only
represent a subset of the meaningful data domain for the chart. In the
dot plot, the data domain for the {\em GPA}~{\tiny
  $\xrightarrow{L}$}~{\em yPos} mapping would initially range from
$[2.1, 3.8]$.  To recover the complete meaningful data domain we
consider all linear mappings to same mark attribute ({\em yPos} in our
example) which also have overlapping min/max data ranges. We assume
that all such mappings share the same data domain and compute the
domain as the union of the overlapping min/max data ranges. In our
example, this approach considers the min/max data range for {\em yPos}
mappings of the dots and of the y-axis axis tick marks together and
thereby recovers the complete $[0.0, 4.0]$ data domain.

%%  LocalWords:  Agrawala SVG Agrawala's attr xPos yPos gridlines
% LocalWords:  rect deconID boolean postfixed postfix

%%  LocalWords:  Agrawala's

\bibliographystyle{abbrv}
%%use following if all content of bibtex file should be shown
%\nocite{*}
\bibliography{styletransfer}

\begin{wrapfigure}[6]{l}{1.1in}
\vspace{-0.3in}
\center{
\includegraphics[width=1in]{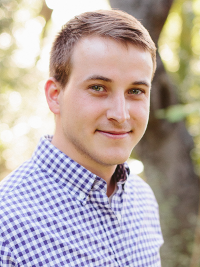}
}
\end{wrapfigure}
\noindent
{\bf \large Jonathan Harper} is a software engineer working at Strava to 
build tools that help athletes better understand and share their
activities.  Previously, he was a graduate student at UC Berkeley
building information visualization design tools.\\[4em]

\begin{wrapfigure}[11]{l}{1.1in}
\vspace{-0.3in}
\center{
\includegraphics[width=1in]{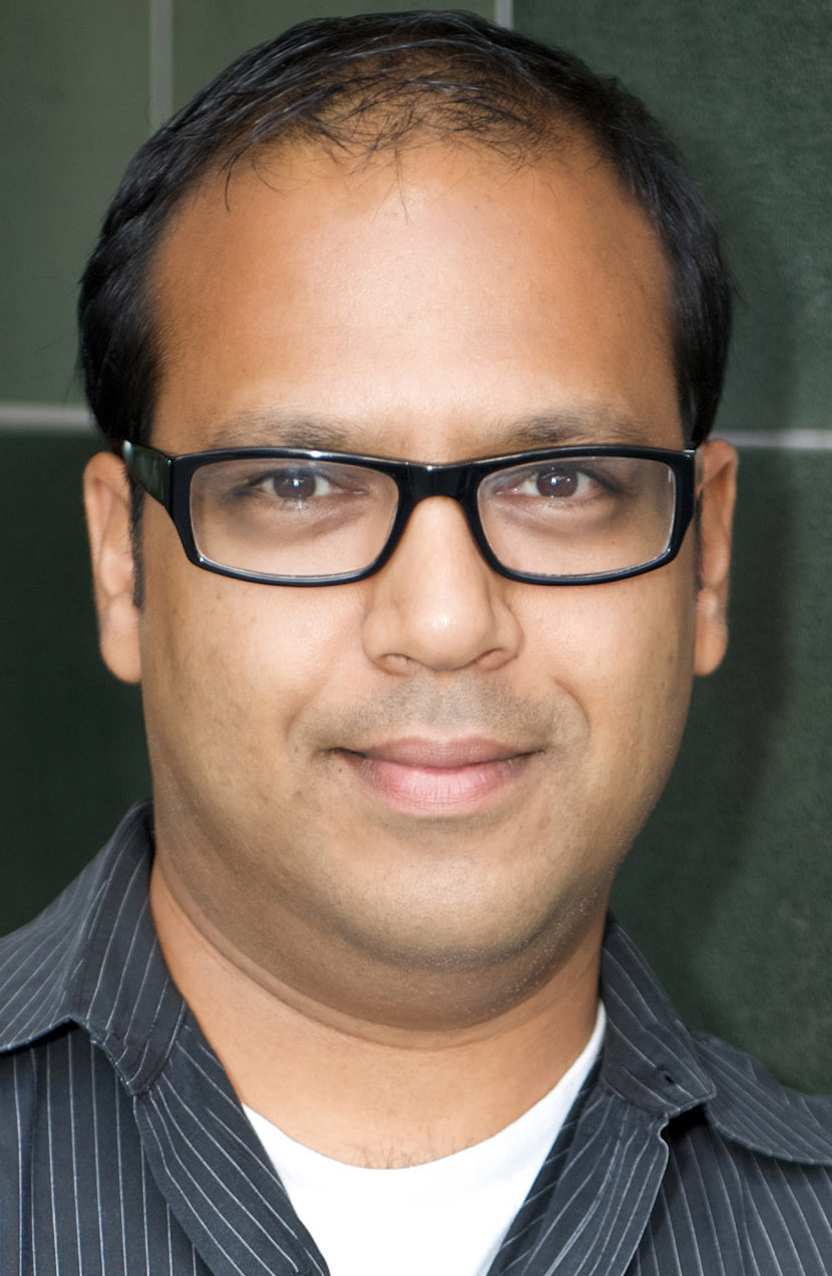}
}
\end{wrapfigure}
\noindent
{\bf \large Maneesh Agrawala} is a Professor of Computer Science and
Director of the Brown Institute for Media Innovation at Stanford
University. He works on computer graphics, human computer interaction
and visualization. His focus is on investigating how cognitive design
principles can be used to improve the effectiveness of audio/visual
media. The goals of this work are to discover the design principles
and then instantiate them in both interactive and automated design
tools.\\

\end{document}